\documentclass[aps,pre,reprint,showpacs,bibnotes,twoside,eqsecnum]{revtex4-1}

\usepackage{amsmath}
\usepackage{amssymb}
\usepackage{epsfig}
\usepackage{bm}
\usepackage{latexsym}
\usepackage{color}

\begin{document}

\title{Resonant enhancement of Anderson localization: Analytical approach}

\author{I.~F.~Herrera-Gonz\'alez}
\affiliation{Instituto de F\'{\i}sica, Benem\'erita Universidad Aut\'{o}noma de Puebla, \\Apartado Postal J-48, Puebla, Pue., 72570, M\'{e}xico}

\author{F.~M.~Izrailev}
\email{izrailev@sirio.ifuap.buap.mx}
\affiliation{Instituto de F\'{\i}sica, Benem\'erita Universidad Aut\'{o}noma de Puebla, \\Apartado Postal J-48, Puebla, Pue., 72570, M\'{e}xico}

\author{N.~M.~Makarov}
\email{nykolay.makarov@correo.buap.mx}
\affiliation{Instituto de Ciencias, Benem\'erita Universidad Aut\'{o}noma de Puebla, \\Priv. 17 Norte No. 3417, Col. San Miguel Hueyotlipan, Puebla, Pue., 72050, M\'{e}xico}


\date{\today}

\begin{abstract}
We study localization properties of the eigenstates and wave transport in one-dimensional system consisting of a set of barriers/wells of fixed thickness and random heights. The inherent peculiarity of the system resulting in the enhanced Anderson localization, is the presence of the resonances emerging due to the coherent interaction of the waves reflected from the interfaces between wells/barriers. Our theoretical approach allows to derive the localization length in infinite samples both out of the resonances and close to them. We examine how the transport properties of finite samples can be described in terms of this length. It is shown that the analytical expressions obtained by standard methods for continuous random potentials can be used in our discrete model, in spite of the presence of resonances that cannot be described by conventional theories. We also discuss whether the single parameter scaling is valid in view of the suggested modification of the theory. All our results are illustrated with numerical data manifesting an excellent agreement with the theory.
\end{abstract}


\maketitle

\section{Introduction}

To date, the theory of Anderson localization in one-dimensional disordered systems is developed in great detail. In particular, various analytical approaches to the models with continuous potentials allow to derive all transport characteristics in dependence on the disorder strength and size of the samples (see, for example, Refs.~\cite{LGP88,IKM12}). On the other hand, for the tight-binding and Kronig-Penney models the rigorous analysis is a difficult task due to the presence of resonances of the Fabry-Perot type. The famous example is the standard tight-binding Anderson model for which at the band center the correct expressions for the localization length $L_{\mathrm{loc}}$ and transmission coefficient $T$ can be obtained with specific methods only \cite{KW81,DG84,GNS94,RT98,ST03,ST03a,DIK04,THI12}. As was found, the band center corresponds to the lowest (most important) resonance resulting in a non-flat distribution $\rho(\theta)$ of the phase $\theta$ of wave function, emerging when the wave propagates along a disordered sample. The same situation occurs for Kronig-Penney models with weak disorder for which the Fabry-Perot resonances have to be taken into account, if the aim is to develop general expressions valid for any value of energy inside allowed energy bands. Although away from these resonances the analytical results for $L_{\mathrm{loc}}$  and $T$ can be obtained relatively easy, in the vicinity of the resonances the transport properties are mainly understood with the use of numerical simulations.

One of the open problems, in connection with the presence of these resonances, is how to relate global transmission characteristics to the localization length $L_{\mathrm{loc}}$ which near the resonances can be obtained with one of specific methods. In contrast with continuous scattering potentials for which the so-called single parameter scaling (SPS) holds, the question about the validity of the SPS for tight-binding and Kronig-Penney models remains open. In the theory of scattering for continuous one-dimensional models the SPS is trivially valid since the distribution of $T$ depends on one single parameter only, which is the ratio between the localization length $L_{\mathrm{loc}}$ and size of the sample $L$. This means that the knowledge of the localization length (defined in the limit $L\to\infty$) gives a complete solution of the scattering problem. As for the tight-binding and Kronig-Penney models, the relation between the localization length and transport properties in the regions close to resonances is typically unknown, the fact that makes the SPS hypothesis questionable. Therefore, for such systems the problem of the relevance of $L_{\mathrm{loc}}$ to transport properties is of great importance.

Recently, the detailed study of the transmittance and reflectance in the vicinity of the lowest resonance has been performed in Ref.~\cite{DMYT12}. The authors where able to develop the theory and obtain the analytical results for a quite simple model for which the potential consists of barriers and wells of a fixed thickness $d$, however, with a weak variation of their heights and depths. The lowest resonance emerges when the phase shift $\varphi$ of the wave equals $\pi$ after passing freely a barrier or well. Numerical simulation shows that for such a value of $\varphi$ a clear dip occurs for the transmission coefficient. It was shown how to describe the Landauer resistance and transmittance in the ballistic regime, with the use of special technics based on the so-called ``building block" method.

In our paper we analyze the same model, however, paying main attention to the relevance of the localization length to transport characteristics, namely, to the transmission coefficient $T$, its logarithm $\ln T$, and their variances. To do this, we develop a new method consisting of few steps. First, we show how to find an analytical expression for the phase distribution $\rho(\theta)$ which is highly non-uniform near the resonances. Then, with this distribution we demonstrate how the  localization length can be analytically obtained in the vicinity of any resonance $\varphi=j\pi$
with $j$ integer. Finally, we show that if to insert this localization length into the expression for the moments of $T$ obtained for the models with random continuous potentials, one gets a nice correspondence with numerical data in a whole energy region including the resonances. The comparison with numerical data is excellent outside the ballistic regime, i.e. in the region of a strong and intermediate localization, when the obtained localization length $L_{\mathrm{loc}}$ is smaller than or of the order of the system size $L$. We also suggest how to improve the correspondence for the ballistic regime, indicating that the localization length found in a strong limit $L\to\infty$ may have no physical sense since for a not large enough value of $L$ the phase distribution $\rho(\theta)$ is still non-stationary. To overcame this problem we suggested to use the finite-length Lyapunov exponent which can be computed numerically. Then, the data show much better agreement with this semi-analytical approach.

\section{Model Formulation}

\begin{figure}[t]
\includegraphics[width=8cm,height=5cm]{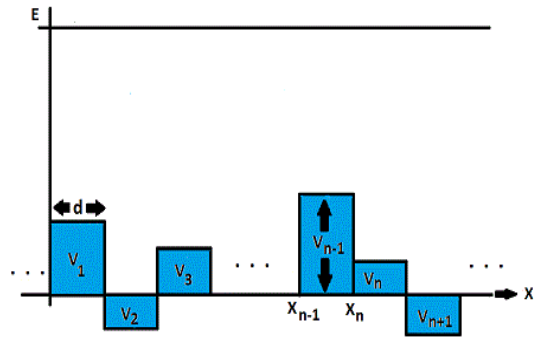}
\caption{\label{PRE-Fig01} (Color online) Set of random barriers and wells.}
\end{figure}

We study the localization and transport properties of a quantum particle with the mass $m$ propagating through an array of rectangular potential barriers and/or wells. The height $V_n$ of the $n$-th scatterer randomly depends on index $n$, however, all the barriers/wells are of the same thickness $d$ (see Fig.~\ref{PRE-Fig01}). Our study is restricted to the case of the over-barrier scattering when the particle energy $E$ is much larger than the strength of the random potential, $\langle V_n^2\rangle\ll E^2$. The stationary Schr\"odinger equation for the wave function $\psi_n(x)$ of the particle traveling over the $n$-th barrier/well, reads
\begin{equation}\label{eq:Schr-eq}
\left(\frac{d^2}{dx^2}+k^2_n\right)\psi_n(x)=0,
\end{equation}
where $k_n=\sqrt{2m\left(E-V_n\right)/\hbar^2}$ is the wave number of the particle. Its general solution can be presented as a superposition of two standing waves,
\begin{eqnarray}\label{eq:Psi}
\psi_{n}(x)=\psi_{n}(x_n)\cos\left[k_{n}(x-x_{n})\right]\\
+k_{n}^{-1}\psi'_{n}(x_{n})\sin\left[k_{n}(x-x_{n})\right],\nonumber
\end{eqnarray}
for $x_{n}\leqslant x\leqslant x_{n+1}$. The $x$-axis is directed along the array with $x=x_{n}$ standing for the coordinate of the left-hand edge of the $n$-th barrier, see Fig.~\ref{PRE-Fig01}. The prime implies the derivative with respect to $x$. Note that the constant thickness of the unit $n$-th barrier is defined as
\begin{equation}\label{eq:d-def}
d=x_{n+1}-x_{n}.
\end{equation}

The general solution \eqref{eq:Psi} has to be complemented by two continuity conditions at the interfaces between neighboring barriers/wells,
\begin{equation}\label{eq:BC}
\begin{split}
&\psi_{n}(x_{n+1})=\psi_{n+1}(x_{n+1}),\\
&\psi'_{n}(x_{n+1})=\psi'_{n+1}(x_{n+1}).
\end{split}
\end{equation}
The combination of Eqs.~\eqref{eq:Psi} and \eqref{eq:BC} yields the recurrent relations describing the wave-function transfer through the $n$-th unit cell of the array,
\begin{equation}\label{eq:Ham-map}
\begin{split}
&Q_{n+1}=Q_n\cos\varphi_n+P_n\frac{\varphi}{\varphi_n}\sin\varphi_n,\\
&P_{n+1}=-Q_n\frac{\varphi_n}{\varphi}\sin\varphi_n+P_n\cos\varphi_n.
\end{split}
\end{equation}
Here $Q_n$ and $P_n$ refer to the wave-function and its re\-scaled derivative, respectively, taken at left-hand edge of the $n$-th unit cell,
\begin{equation}\label{eq:QP-def}
Q_n=\psi_n(x_n),\qquad P_n=k^{-1}\psi'_n(x_n).
\end{equation}
The phase shift $\varphi_n$ randomized by the compositional disorder, and its unperturbed counterpart $\varphi$ are defined by
\begin{equation}\label{eq:phi-def}
\varphi_n=k_nd=\varphi\sqrt{1-v_n/\varphi^2}\,,\qquad\varphi=kd\,.
\end{equation}
Here, for convenience, we have introduced the dimensionless random strength $v_n$ of the potential barriers/wells and the unperturbed particle wave number $k$,
\begin{equation}\label{eq:vn-k-def}
v_n=2mV_nd^2/\hbar^2,\qquad k=\sqrt{2mE/\hbar^2}.
\end{equation}

Remarkably, the recurrent relations \eqref{eq:Ham-map} can be regarded as the classical \emph{Hamiltonian map} describing the evolution of trajectories in the phase space $(Q,P)$ with discrete time $n$ for a linear oscillator subjected to the time-dependent parametric force. In such a representation $Q_n$ and $P_n$ can be treated as the classical coordinate and momentum, respectively \cite{IKT95}. Thus, the problem of quantum localization can be formally reduced to the analysis of the energetic instability of a stochastic oscillator \cite{TI00}. Note that the Hamiltonian map \eqref{eq:Ham-map} belongs to the class of area-preserving maps whose determinant equals unity.

For the analytical study it is convenient to pass to polar coordinates, namely, to the radius $R_n$ and angle $\theta_n$,
\begin{equation}\label{eq:QP-RTheta}
Q_n=R_n\cos\theta_n,\qquad P_n=R_n\sin\theta_n.
\end{equation}
According to Eq.~\eqref{eq:Ham-map}, the Hamiltonian map in the \emph{radius-angle presentation} gets the form,
\begin{eqnarray}
\frac{R^2_{n+1}}{R^2_n}&=&\cos^2\varphi_n+\frac{1}{2}\left(\frac{\varphi}{\varphi_n}-
\frac{\varphi_n}{\varphi}\right)\sin2\varphi_n\sin2\theta_n\nonumber\\
&+&\left(\frac{\varphi^2}{\varphi^2_n}\sin^2\theta_n+\frac{\varphi^2_n}{\varphi^2}\cos^2\theta_n\right)\sin^2\varphi_n,\label{eq:map-R}\\[6pt]
\tan\theta_{n+1}&=&\frac{-(\varphi_n/\varphi)\sin\varphi_n+\cos\varphi_n\tan\theta_n}
{\cos\varphi_n+(\varphi/\varphi_n)\sin\varphi_n\tan\theta_n}.\label{eq:map-theta}
\end{eqnarray}
As one can see, the {\it linear} two-dimensional map \eqref{eq:Ham-map} for $Q_n$ and $P_n$ can be reduced to the {\it non-linear} one-dimensional map for the angle $\theta_n$ only. It should be stressed that in this map the angle $\theta_n$ can be considered in the range $[0,2\pi]$. Note also that both maps are the time dependent ones, this fact makes the rigorous analysis problematic.

In line with the concept of the Hamiltonian map, the localization length $L_{\mathrm{loc}}$ is determined by the rate of exponential growth of the coordinate $Q_n$ or momentum $P_n$, once the initial conditions ($Q_0,P_0$) are specified. The conventional definition of the localization length $L_{\mathrm{loc}}$ is due to the inverse of the Lyapunov exponent $\lambda$, and the latter can be defined as
\begin{equation}\label{eq:Lloc-Lyap-def01}
\frac{d}{L_{\mathrm{loc}}}\equiv\lambda=\lim_{N\to\infty}\frac{1}{N}
\sum_{n=1}^N\ln\left|\frac{Q_{n+1}}{Q_n}\right|.
\end{equation}
Another definition which gives the same result, takes the form~\cite{IKM12,IKT95},
\begin{equation}\label{eq:Lloc-Lyap-def02}
\lambda= \frac{1}{2}\Big\langle\ln\frac{R^2_{n+1}}{R^2_n}\Big\rangle=
-\frac{1}{2}\Big\langle\ln\frac{d\theta_{n+1}}{d\theta_{n}}\Big\rangle.
\end{equation}
Here the averaging $\langle ... \rangle$ is performed along the discrete ``time" $n$. The second relation in \eqref{eq:Lloc-Lyap-def02} can be derived directly from Eqs.~\eqref{eq:map-R} -- \eqref{eq:map-theta}. It is useful for the analytical analysis since due to ergodicity the averaging over $n$
can be substituted by the averaging over different realizations of the disorder $v_n$ and random phase $\theta_n$.

In what follows, the dimensionless variable $v_n$ imposing the disorder, is specified by a random sequence of white-noise type with the zero average and variance $\sigma^2$,
\begin{equation}\label{eq:vn-corr}
\langle v_n\rangle=0,\quad\langle v_n^2\rangle=\sigma^2,\quad\langle v_{n}v_{n'}\rangle=\sigma^2\delta_{nn'}.
\end{equation}
In numerical analysis, when generating random sequence $v_n$ we use the entries of the uniform box probability distribution inside a finite interval $[-w,w]$ with   the variance $\sigma^2=w^2/3$. However, our analytical results are valid for any distribution of $v_n$ with correlation properties \eqref{eq:vn-corr} and finite small variance.

In terms of the statistical characteristics \eqref{eq:vn-corr} for the random quantity $v_n$ the conditions of weak disorder
($v_n\ll\varphi$, $v_n\ll\varphi^2$) and the over-barrier scattering ($v_n<\varphi^2$) can be rewritten in the explicit form,
\begin{equation}\label{eq:WeakDis}
\sigma^2\ll\varphi^2,\qquad\sigma^2\ll\varphi^4.
\end{equation}
These conditions allow us to develop a proper perturbation theory.

It is worthwhile to mention that the system under consideration is similar to an array of optic slabs with random and frequency dispersive refractive indices. The features of optic wave localization in the non-dispersive array, where the refractive index is independent of the wave frequency, were analyzed in detail in Refs.~\cite{IKM12,IMT10,TIM11,TIM12,BO12,TIM13,Go12}. As one can recognize below, the effect of energy/frequency dispersion drastically changes the localization properties of both quantum and optic disordered systems.

\section{Non-Resonant Localization Length}

For a weak disorder, see \eqref{eq:WeakDis}, we expand the $R$-map \eqref{eq:map-R} up to the second order in perturbation $v_n$. Then, we substitute the result into Eq.~\eqref{eq:Lloc-Lyap-def02} with the subsequent expansion of the logarithm, keeping the terms quadratic in disorder. Taking into account that we consider the case of a white-noise disorder, one can neglect high-order correlations between the disorder $v_n$ and phase $\theta_n$ \cite{LGP88}. This allows us to perform the statistical averaging over $v_n$ in accordance with the correlation properties \eqref{eq:vn-corr}. As a
result, we arrive at the following quadratic approximation for the Lyapunov exponent,
\begin{eqnarray}\label{eq:Lyap-appr}
\lambda&=&\frac{\sigma^2}{8\varphi^4}\Big[\sin^2\varphi+\overline{\sin2\theta}\left(\frac{3\sin2\varphi}{4}-2\varphi\cos2\varphi\right)\nonumber\\
&-&\overline{\cos2\theta}\left(\frac{3\sin^2\varphi}{2}-2\varphi\sin2\varphi\right)\\
&+&\overline{\sin4\theta}\sin^2\varphi\sin2\varphi+\overline{\cos4\theta}\sin^2\varphi\cos2\varphi\Big].\nonumber
\end{eqnarray}
Here we substituted the averaging of $\theta_n$ over $n$ by the statistical average over $\theta$ (denoted by the bar) assuming that the distribution $\rho(\theta)$ exists. The starting point for obtaining $\rho(\theta)$ is the quadratic expansion of the $\theta$-map \eqref{eq:map-theta},
\begin{eqnarray}\label{eq:thetaMap-appr}
\theta_{n+1}-\theta_n=-\varphi+\frac{v_n}{2\varphi^2}\big[\varphi+\sin\varphi\cos(2\theta_n-\varphi)\big]\qquad\\
+\frac{v_n^2}{8\varphi^4}\big[\varphi+\sin \varphi\cos(2\theta_n-\varphi)-2\varphi\cos(2\theta_n-2\varphi)\nonumber\\
-\sin^2\varphi\sin(4\theta_n-2\varphi)-2\sin\theta_n\sin\varphi\sin(\theta_n-\varphi)\big].\nonumber
\end{eqnarray}

By analyzing Eq.~\eqref{eq:thetaMap-appr} one can suggest that for non-zero values of $\varphi$ only a small number of iterations are needed for $\theta_n$ to fill the whole interval [$0,2\pi$]. Therefore, in this case a uniform phase distribution can be expected in the lowest order of perturbation,
\begin{equation}\label{eq:RhoFlat}
\rho(\theta)=1/2\pi.
\end{equation}
The averaging of Eq.~\eqref{eq:Lyap-appr} with the probability density \eqref{eq:RhoFlat} is trivial and gives rise to the expression
\begin{equation}\label{eq:Lyap-flat}
\lambda=\frac{\sigma^2}{8\varphi^4}\sin^2\varphi.
\end{equation}
This result is in a complete correspondence with those previously obtained for discrete optic systems with randomized refractive index
(see, e.g., Refs.~\cite{IMT10,TIM11,TIM12,TIM13}).

However, the flat distribution \eqref{eq:RhoFlat} may not be valid for the resonant values of $\varphi$, namely, for $\varphi=2\pi r/q$ with $r, q$ integers. For such rational values (with respect to $2\pi$) the unperturbed trajectory $\theta_n$ is the periodic orbit with the period $q$, therefore, $\rho (\theta)$ is the periodic delta-function of the same period. By adding a weak disorder the phases $\theta_n$ begin to diffuse around each of the delta-peaks, and it is not clear whether the fingerprint of these periodic orbits disappears in the form of $\rho(\theta)$ in the limit $n \rightarrow \infty$. One can expect that the strongest resonances correspond to $q=1$ and $q=2$. Below we restrict our study by these resonances only. As for high-order resonances with $q>2$, it is quite naturally to expect that they give much less influence to the localization length, if any.

\begin{figure}[t]
\includegraphics[width=8cm,height=5cm]{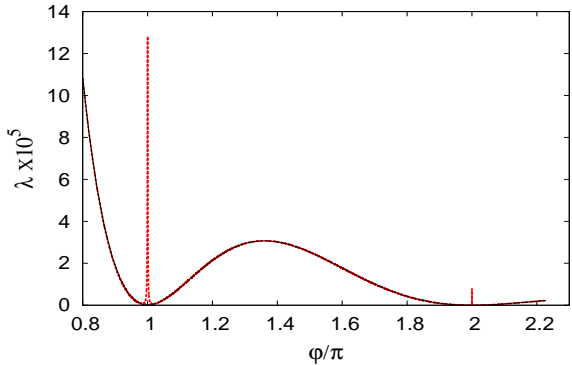}
\caption{\label{PRE-Fig02} (Color online) Lyapunov exponent $\lambda$ vs phase shift $\varphi$: Continuous curve depicts the analytical result \eqref{eq:Lyap-flat}, dashed curve shows the numerical computation of $\lambda$ obtained with the use of Eq.~\eqref{eq:Lloc-Lyap-def01}. The intensity of disorder is $\sigma^2=0.1$.}
\end{figure}

Indeed, as is displayed in Fig.~\ref{PRE-Fig02}, the Lyapunov exponent \eqref{eq:Lloc-Lyap-def01} obtained from the numerical iteration of the Hamiltonian map \eqref{eq:Ham-map} differs from Eq.~\eqref{eq:Lyap-flat} only very close to the points $\varphi=2j\pi$ and $\varphi=(2j-1)\pi$ ($j$ is an integer), i.e., when $q=1$ and $q=2$. Note that for $\varphi=2j\pi$ the unperturbed $\theta$-map is given by a single point, and for $\varphi=(2j-1)\pi$ it results in two fixed points. Therefore, in the analytical approach one has to treat these two cases separately.

From the physical point of view, the origin of the above peculiarities is the resonance effect emerging due to the coherent interaction of the waves reflected from the boundaries of wells/barriers. This effect may be compared with the well known Fabry-Perot resonances emerging due to multiple reflections of the wave from the interfaces in multi-layered photonic structures. As was shown both theoretically and in experiments~\cite{IKM12,IM09,Lo09,LM09,Do11}, in the non-dispersive systems these resonances strongly suppress the localization. Thus,
the Fabry-Perot resonances are typically associated with the resonance {\it enhancement} of the transmission. In our case, as is seen in Fig.~\ref{PRE-Fig02}, the resonances result in the {\it suppression} of the transmission for both odd and even values of $q$. Such a peculiarity of the lowest resonance $\varphi = \pi$ occurring in the model has been predicted in Ref.~\cite{DMYT12}.

Below the analysis of the resonance effects will be performed in a way similar to that used in the study of peculiarities of the localization arising in the one-dimensional Anderson model near the band center and band adges~\cite{THI12,IKM12}.

\section{Even Resonances}

Here we consider the energy region around the even resonances,
\begin{equation}\label{eq:EvenRes-def}
\varphi=2j\pi+\epsilon,\qquad|\epsilon|\ll1,\qquad j=1,2,3,\ldots,
\end{equation}
were the unperturbed $\theta$-map has almost-periodic orbits of the period one. Therefore, the point $\theta_{n+1}$ is very close to $\theta_n$ provided a weak disorder is imposed. The disordered $\theta$-map near the even resonances \eqref{eq:EvenRes-def} can be obtained from Eq.~\eqref{eq:thetaMap-appr} by the corresponding first-order expansion with respect to a small resonance detuning $\epsilon$. After omitting the term that is integer multiple of $2\pi$, the recurrence relation for the polar angle $\theta$ reads
\begin{equation}\label{eq:EvenRes-thetaMap}
\theta_{n+1}-\theta_n=-\epsilon+\frac{v_n}{4j\pi}+\frac{v^2_n}{64j^3\pi^3}(1-2\cos2\theta_n).
\end{equation}
Here we have neglected the terms containing $v_n\epsilon$ since they do not contribute to the associated Fokker-Plank equation derived within the linear approximation in the detuning $\epsilon$ and quadratic one in the disorder $v_n$.

In order to obtain the phase distribution, one has to derive the stationary Fokker-Plank equation for $\rho(\theta)$. This can be done in the same way as described, e.g., in Refs.~\cite{IKT95,TIM11,TIM12,TIM13,THI12,G04,IRT98,HIT08}. Specifically, we rewrite the map \eqref{eq:EvenRes-thetaMap} in the continuum limit and after replace the random variable $v(t)$ with the Wiener process $W$ in accordance with the definition, $\sigma dW=v(t)dt$. As a result, we come to the so-called It\^o stochastic differential equation,
\begin{equation}\label{eq:EvenRes-Ito}
d\theta=\frac{\sigma}{4j\pi}dW+\left[-\epsilon+\frac{\sigma^2(1-2\cos2\theta)}
{64j^3\pi^3}\right]dt.
\end{equation}
With this equation one can study the dynamics of the stochastic process $\theta(t)$ once the initial condition $\theta(t_0)=\theta_0$ is known. On the other hand, following the theory of stochastic differential equations~\cite{G04}, one can readily associate the It\^o equation \eqref{eq:EvenRes-Ito} with the stationary Fokker-Plank equation for the probability density $\rho_s(\theta)$,
\begin{equation}\label{eq:EvenRes-FP}
\frac{\sigma^2}{32j^2\pi^2}\frac{d^2\rho_s}{d\theta^2}+\frac{d}{d\theta}\left[\epsilon-\frac{\sigma^2(1-2\cos2\theta)}{64j^3\pi^3}\right]\rho_s=0.
\end{equation}
This equation should be complemented by the condition of periodicity and by the normalization condition,
\begin{equation}\label{eq:FP-Cond}
\rho_s(\theta+2\pi)=\rho_s(\theta),\qquad\qquad\int_{0}^{2\pi}\rho_s(\theta)d\theta=1.
\end{equation}
%
\begin{figure}[t]
\includegraphics[width=8cm,height=5cm]{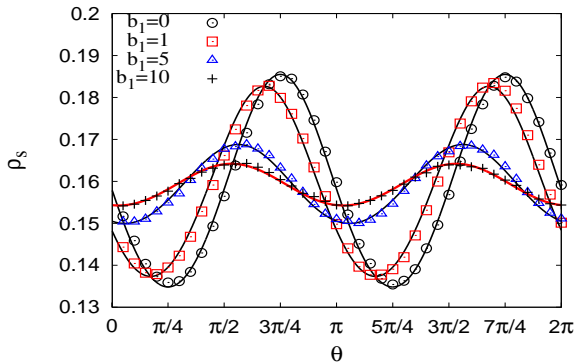}
\caption{\label{PRE-Fig03} (Color online) Distribution $\rho_s(\theta)$ for the even resonance ($j=1$) and various values of $b_1$: Continuous curves show the analytical equation \eqref{eq:EvenRes-rhoMod}, while different symbols correspond to numerical data. The intensity of disorder is $\sigma^2=0.1$.}
\end{figure}
%
After the integration of Eq.~\eqref{eq:EvenRes-FP} we obtain the following linear first-order equation,
\begin{equation}\label{eq:EvenRes-FP1or}
\frac{d\rho_s(\theta)}{d\theta}+\left[b_j+\frac{1}{j\pi}\cos2\theta\right]\rho_s(\theta)=C,
\end{equation}
with the constant $C$ which can be found from the periodicity condition. Remarkably, in Eq.~\eqref{eq:EvenRes-FP1or} the resonance detuning turns out to be modified as follows,
\begin{eqnarray}\label{eq:EvenRes-bj}
b_j&=&\left(\epsilon-\frac{\sigma^2}{64j^3\pi^3}\right)\frac{32j^2\pi^2}{\sigma^2}\nonumber\\
&=&\left(\varphi-2j\pi-\frac{\sigma^2}{64j^3\pi^3}\right)\frac{32j^2\pi^2}{\sigma^2}.
\end{eqnarray}
This means that the even resonances are shifted by disorder to the right,
\begin{equation}\label{eq:EvenRes-ResShift}
\varphi_{\mathrm{res}}=2j\pi+\frac{\sigma^2}{64j^3\pi^3}.
\end{equation}
Also, Eq.~\eqref{eq:EvenRes-bj} manifests an emergence of the disorder-induced scale $\sim\sigma^2/j^2\pi^2$ for the resonance detuning.

\begin{figure}[t]
\includegraphics[width=8cm,height=5cm]{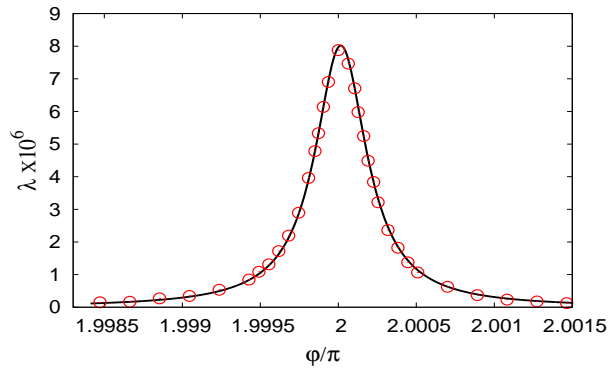}
\caption{\label{PRE-Fig04} (Color online) Rescaled Lyapunov exponent vs phase shift $\varphi$ in the vicinity of the even resonance \eqref{eq:EvenRes-def} with $j=1$ and $\sigma^2=0.1$. Continuous curve corresponds to the analytical result \eqref{eq:EvenRes-Lyap}, circles represent the numerical data.}
\end{figure}

Solving equation \eqref{eq:EvenRes-FP1or} with the periodicity condition \eqref{eq:FP-Cond} yields the nonuniform phase distribution in the vicinity of the even $j$-resonance \eqref{eq:EvenRes-def},
\begin{eqnarray}\label{eq:EvenRes-rho}
\frac{\rho_s(\theta)}{\rho_s(0)}&=&\mbox{e}^{-\mu(\theta)}\left[1+\frac{\mbox{e}^{\mu(2\pi)}-1}{\int^{2\pi}_0\mbox{e}^{\mu(\theta^{'})}d\theta^{'}} \int^{\theta}_0\mbox{e}^{\mu(\theta')}d\theta^{'}\right],\qquad\\
\mu(\theta)&=&\left[b_j\theta +\frac{1}{2j\pi}\sin 2\theta\right].\nonumber
\end{eqnarray}
Here the initial value $\rho_s(0)$ is specified by the normalization condition from Eqs.~\eqref{eq:FP-Cond}.

It is important to emphasize that the initial Eq.~\eqref{eq:Lyap-appr} for the Lyapunov exponent contains only zero, second and fourth harmonics with respect to the $\theta$-phase. Hence, in order to perform the averaging procedure only the corresponding zero, second and fourth harmonics of the
probability density $\rho_s(\theta)$ are needed. All the others give zero result after the averaging. The simplest way to extract the important harmonics from Eq.~\eqref{eq:EvenRes-rho} is to replace the smooth factors in $\exp[-\mu(\theta)]$ and $\exp[\mu(\theta')]$ with their approximate expansions, e.g.,
\begin{equation}\label{eq:EvenRes-smooth}
\exp\left(\frac{\sin2\theta}{2j\pi}\right)=1+\frac{\sin2\theta}{2\pi j}+\frac{\sin^22\theta}{8\pi^2j^2}\,.
\end{equation}
Then, the integrals in Eq.~\eqref{eq:EvenRes-rho} can be taken explicitly that yields the truncated distribution function $\rho_s(\theta)$ as a superposition of the uniform one \eqref{eq:RhoFlat} with oscillating modulations,
\begin{eqnarray}\label{eq:EvenRes-rhoMod}
\rho_s(\theta)&=&\frac{1}{2\pi}-\frac{2\sin2\theta+b_j\cos2\theta}{2j\pi^2(4+b_j^2)}\nonumber\\
&&+\frac{6b_j\sin4\theta+(b_j^2-8)\cos4\theta}{4j^2\pi^3(4+b_j^2)(16+b_j^2)}.
\end{eqnarray}
One can see that at the resonance $\varphi=\varphi_{\mathrm{res}}$ ($b_j=0$) the distribution profile has strong oscillations. However, when the phase shift $\varphi$ moves away from the resonance ($b_j\to\infty$), the oscillations are decreasing, with a smooth (power-law) convergence to the flat distribution. The similar effect takes place as the order of the resonance $j$ increases: The higher the order $j$ the closer the $\theta$-distribution to the flat one. Therefore, the resonances with large $j$ are hardly observable. Fig.~\ref{PRE-Fig03} displays the change of the $\theta$-distribution with the variation of the modified detuning parameter $b_1$ in the vicinity of the even resonance $\varphi\approx2\pi$. The data shown in Fig.~\ref{PRE-Fig03} demonstrate an excellent agreement between the analytical result \eqref{eq:EvenRes-rhoMod} and the corresponding numerical simulation.

Now we are able to perform the averaging in Eq.~\eqref{eq:Lyap-appr} with the use of the probability density \eqref{eq:EvenRes-rhoMod}. Within the lowest approximation in disorder $\sigma^2$ and in resonance detuning $\epsilon$, the non-zero contribution comes from the first and second terms in both Eqs.~\eqref{eq:Lyap-appr} and \eqref{eq:EvenRes-rhoMod}. As a result we get
\begin{equation}\label{eq:EvenRes-Lyap}
\lambda=\frac{\sigma^2}{8\varphi^4}\left(\sin^2\varphi+\frac{4}{4+b^2_j}\right).
\end{equation}
Fig.~\ref{PRE-Fig04} compares this expression with the numerical data for the Lyapunov exponent in the vicinity of the first ($j=1$) even resonance.

\section{Odd Resonances}

The energy region of the odd resonances is defined by condition
\begin{equation}\label{eq:OddRes-def}
\varphi=(2j-1)\pi+\epsilon,\qquad|\epsilon|\ll1,\qquad j=1,2,3,\ldots.
\end{equation}
At odd resonances the unperturbed $\theta$-map for a fixed initial value $\theta_0$ has periodic orbits with period 2 and is presented by two points. Therefore, the phase $\theta_{n+2}$ coincides with $\theta_n$. For a weak disorder, due to a small resonance detuning $\epsilon$ these points are no more fixed, however, after two steps the distance between them is quite small. For this reason it is convenient to treat the two-step recurrent $\theta$-relation between $\theta_{n+2}$ and $\theta_n$. This relation is readily obtained by the iteration of the general map \eqref{eq:thetaMap-appr}. Within the lowest order in the resonance detuning $\epsilon$ and with the use of statistical independence of the variables $v_n$ and $v_{n+1}$, see Eq.~\eqref{eq:vn-corr}, one gets
\begin{eqnarray}\label{eq:OddRes-thetaMap}
\theta_{n+2}-\theta_n&=&-2\epsilon+\frac{v_n+v_{n+1}}{2(2j-1)\pi}\nonumber\\
&&+\frac{v_n^2+v_{n+1}^2}{8(2j-1)^3\pi^3}(1-2\cos 2\theta_n).
\end{eqnarray}

Being written in the continuum limit in the terms of two independent Wiener processes, $W_1$ and $W_2$, the $\theta$-map~\eqref{eq:OddRes-thetaMap} takes the form of the It\^o equation,
\begin{eqnarray}\label{eq:OddRes-Ito}
d\theta&=&\frac{\sigma}{2(2j-1)\pi}(dW_1+dW_2)\nonumber\\
&&+\left[-2\epsilon+\frac{\sigma^2(1-2\cos2\theta)}{4(2j-1)^3\pi^3}\right]dt.
\end{eqnarray}
It is important that the random processes $W_1$ and $W_2$ have the same statistical properties. Therefore, as in the previous Section, we can apply the method described in Ref.~\cite{G04} and write down the corresponding Fokker-Plank equation for the stationary distribution function $\rho_s(\theta)$,
\begin{equation}\label{eq:OddRes-FP}
\frac{\sigma^2}{8(2j-1)^2\pi^2}\frac{d^2\rho_s}{d\theta^2}+
\frac{d}{d\theta}\left[\epsilon-\frac{\sigma^2(1-2\cos2\theta)}{8(2j-1)^3\pi^3}\right]\rho_s=0.
\end{equation}
This equation is complemented by the conditions \eqref{eq:FP-Cond} of periodicity and normalization.

\begin{figure}[t]
\includegraphics[width=8cm,height=5cm]{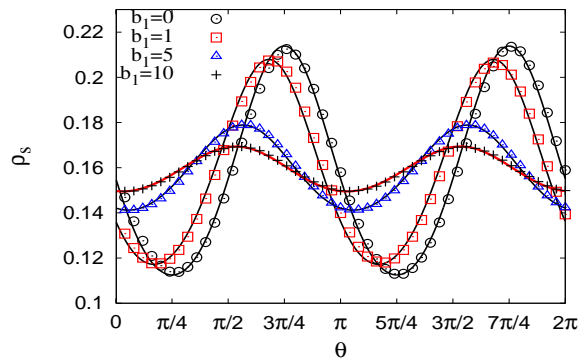}
\caption{\label{PRE-Fig05} (Color online) Distribution $\rho_s(\theta)$ for the first odd resonance ($j=1$) and various $b_1$: Continuous curves show the analytical equation \eqref{eq:OddRes-rhoMod} while different symbols stand for the numerical data. The intensity of disorder is $\sigma^2=0.1$.}
\end{figure}

It is clear that the solution of Eqs.~\eqref{eq:OddRes-FP}, \eqref{eq:FP-Cond} has the form \eqref{eq:EvenRes-rho}, however, with a new function $\mu(\theta)$,
\begin{equation}\label{eq:OddRes-mu}
\mu(\theta)=\left[b_j\theta+\frac{1}{(2j-1)\pi}\sin2\theta\right],
\end{equation}
and with another resonance detuning $b_j$,
\begin{eqnarray}\label{eq:OddRes-bj}
b_j&=&\left[\epsilon-\frac{\sigma^2}{8(2j-1)^3\pi^3}\right]\frac{8(2j-1)^2\pi^2}{\sigma^2}\\
&=&\left[\varphi-(2j-1)\pi-\frac{\sigma^2}{8(2j-1)^3\pi^3}\right]\frac{8(2j-1)^2\pi^2}{\sigma^2}.\nonumber
\end{eqnarray}
Thus, the odd resonances are shifted by disorder exactly as even resonances,
\begin{equation}\label{eq:OddRes-ResShift}
\varphi_{\mathrm{res}}=(2j-1)\pi+\frac{\sigma^2}{8(2j-1)^3\pi^3},
\end{equation}
and have similar disorder-induced broadening of the order of $\sigma^2/(2j-1)^2\pi^2$.

The truncated distribution function $\rho_s(\theta)$ containing only zero, second and fourth harmonics that contribute to the averaging
of Eq.~\eqref{eq:Lyap-appr}, can be extracted from the equations \eqref{eq:EvenRes-rho}, \eqref{eq:OddRes-mu}, \eqref{eq:OddRes-bj} in the same manner as for even resonances. As a result, we have
\begin{eqnarray}\label{eq:OddRes-rhoMod}
\rho_s(\theta)&=&\frac{1}{2\pi}-\frac{2\sin2\theta+b_j\cos 2\theta}{(2j-1)\pi^2(4+b^2_j)}\nonumber\\
&&+\frac{6b_j\sin4\theta+(b^2_j-8)\cos4\theta}{(2j-1)^2\pi^3(4+b^2_j)(16+b^2_j)}.
\end{eqnarray}
One can see that this probability density has the structure similar to the truncated distribution \eqref{eq:EvenRes-rhoMod} obtained for even resonances. Specifically, at the resonance $\varphi=\varphi_{\mathrm{res}}$ ($b_j=0$) the distribution \eqref{eq:OddRes-rhoMod} oscillates. These oscillations decrease and smoothly disappear when the phase shift $\varphi$ moves away from the resonance ($b_j\to\infty$). As the order of the odd resonance $j$ increases the oscillations begin to be smoother. Fig.~\ref{PRE-Fig05} shows the change of the $\theta$-distribution with the variation of the modified detuning $b_1$ in the vicinity of the first ($j=1$) odd resonance $\varphi\approx\pi$. Results confirm the validity of the theoretically obtained equation \eqref{eq:OddRes-rhoMod} in comparison with the corresponding numerical data.

After the averaging of Eq.~\eqref{eq:Lyap-appr} with the distribution function \eqref{eq:OddRes-rhoMod} we obtain the Lyapunov exponent for odd resonances \eqref{eq:OddRes-def},
\begin{equation}\label{eq:OddRes-Lyap}
\lambda=\frac{\sigma^2}{8\varphi^4}\left(\sin^2\varphi+\frac{4}{4+b^2_j}\right).
\end{equation}
As one can see, it is actually of the same form as Eq.~\eqref{eq:EvenRes-Lyap} derived for even resonances \eqref{eq:EvenRes-def}. The only difference is in the definition of the modified resonance detuning: for odd resonances $b_j$ contains the quantity $2j-1$ instead of $2j$ for even resonances, compare Eqs.~\eqref{eq:EvenRes-bj} and \eqref{eq:OddRes-bj}.

\section{Interpolated Expression for the Lyapunov Exponent}

Let us now compare expressions \eqref{eq:Lyap-flat}, \eqref{eq:EvenRes-Lyap} and \eqref{eq:OddRes-Lyap} for the Lyapunov exponent that are valid away from resonances and in the vicinity of odd/even resonances, respectively. From this comparison one can easily conclude that Eq.~\eqref{eq:EvenRes-Lyap} or, the same, Eq.~\eqref{eq:OddRes-Lyap} can be regarded as the general interpolation for the Lyapunov exponent if to write the parameter $b_j$ in the generalized form. In order to realize this idea, in Eqs.~\eqref{eq:EvenRes-bj} or \eqref{eq:OddRes-bj} we replace the quantities $2j\pi$ and $(2j-1)\pi$ with the phase shift $\varphi$, and generalize the definition of the resonance detuning $\epsilon$. The explicit result reads
\begin{eqnarray}\label{eq:Lyap-Fin}
&&\lambda(\varphi)=\frac{\sigma^2}{8\varphi^4}\left[\sin^2\varphi+\frac{(\sigma^2/4\varphi^2)^2}{(\sigma^2/4\varphi^2)^2+
(\epsilon-\sigma^2/8\varphi^3)^2}\right];\nonumber\\[6pt]
&&\epsilon=\left\{\begin{array}{ccc}
\varphi-\left[\frac{\varphi}{\pi}\right]\pi&\mbox{for}\quad0\leqslant\frac{\varphi}{\pi}-\left[\frac{\varphi}{\pi}\right]\leqslant\frac{1}{2},\\[6pt]                \varphi-(\left[\frac{\varphi}{\pi}\right]+1)\pi&\mbox{for}\quad\frac{1}{2}<\frac{\varphi}{\pi}-\left[\frac{\varphi}{\pi}\right]<1,\end{array}\right.
\end{eqnarray}
where $[...]$ stands for the integer part. Within the qua\-dratic approximation in disorder, equation \eqref{eq:Lyap-Fin} adequately describes the Lyapunov exponent $\lambda$ inside a wide range of the phase shift $\varphi\propto\sqrt{E}$. The applicability of Eq.~\eqref{eq:Lyap-Fin} is restricted only by the conditions of weak disorder and over-barrier scattering \eqref{eq:WeakDis}.

\begin{figure}[!t]
\includegraphics[width=8cm,height=5cm]{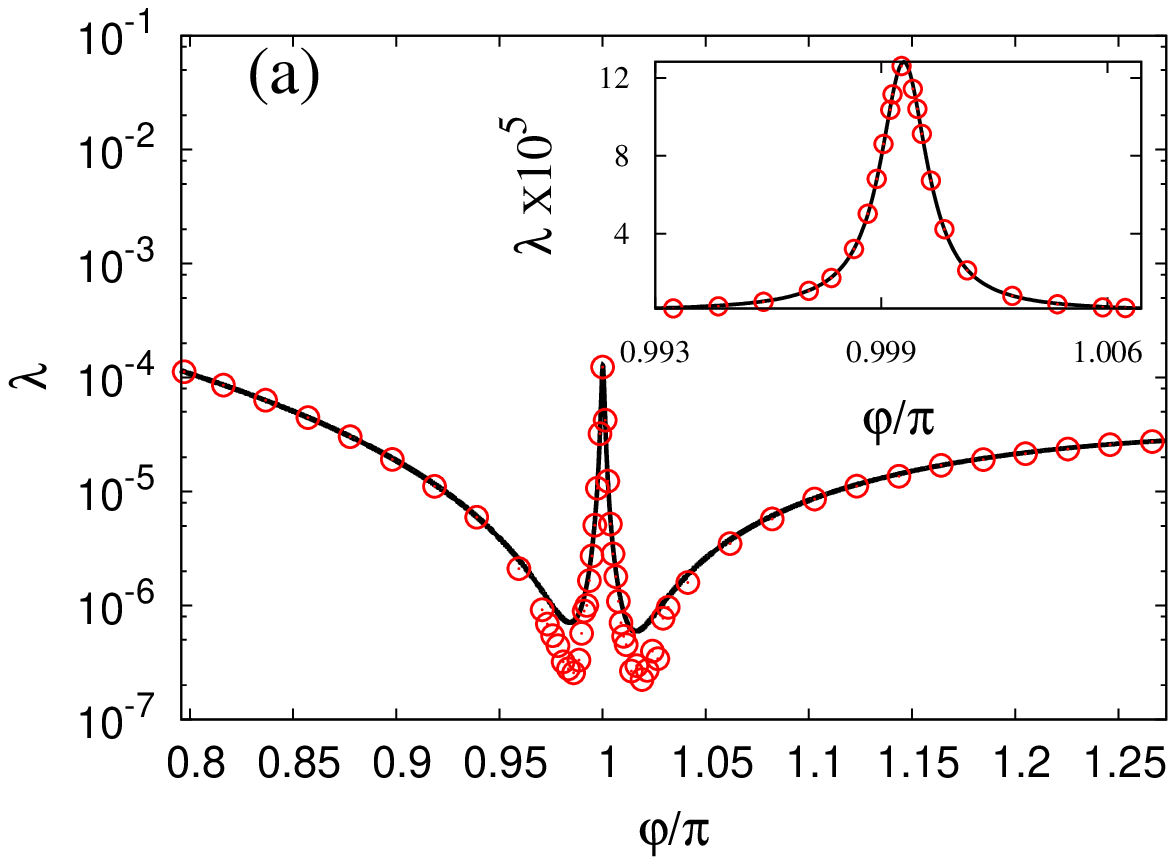}
\includegraphics[width=8cm,height=5cm]{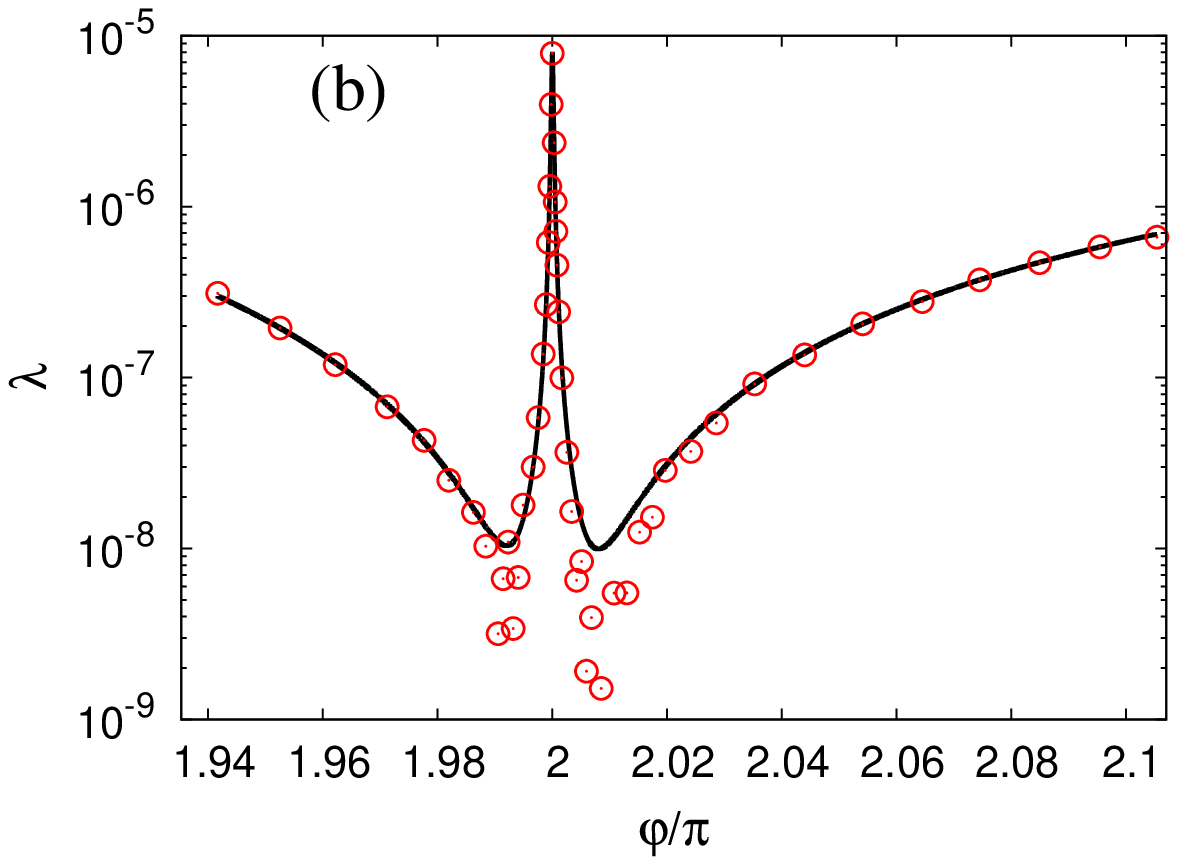}
\caption{\label{PRE-Fig06} (Color online) Resonant line-shape of the Lyapunov exponent for first odd and even resonances, panels (a) and (b), respectively. Continuous curve depicts analytical interpolation \eqref{eq:Lyap-Fin} whereas the circles show corresponding numerical data. The intensity of disorder is $\sigma^2=0.1$.}
\end{figure}

Out of the resonances, the detuning is of the order of unity, $\epsilon\sim1$. The second term in square brackets is negligibly small being of the order of $(\sigma^2/4\varphi^2)^2$. Due to this fact, Eq.~\eqref{eq:Lyap-Fin} is equivalent to Eq.~\eqref{eq:Lyap-flat} in the region between the neighboring resonances.

The second (resonant) term in the square brackets has the Lorentzian form. For both odd and even resonances we have, $\varphi=j\pi+\sigma^2/8\varphi^3$, therefore, this term is equals 1 and strongly prevails over the first term which is of the order of
$(\sigma^2/8\varphi^3)^2$. The half-width of the resonances is $\sigma^2/4\varphi^2$, thus, showing that the resonant line-shape is very sharp. It is worthwhile to note that the higher the resonance order $j$ the sharper the resonance, however, the smaller its amplitude $\sigma^2/8\varphi^4$. When the phase shift $\varphi$ moves away from the resonance the main contribution turns in Eq.~\eqref{eq:Lyap-Fin} from the second term to first one.

Fig.~\ref{PRE-Fig06} shows that the interpolation \eqref{eq:Lyap-Fin} provides a good agreement with the numerical data, apart from the transition regions where the Lyapunov exponent is so small that the perturbation terms of higher order have to be taken into account.

\section{Transport properties}

Now, in connection with the localization length we address the problem of transport properties in finite systems constructed by an array of $N$ unit cells (random barriers and/or wells). In terms of trajectories of the Hamiltonian map \eqref{eq:Ham-map}, \eqref{eq:QP-RTheta} the transmittance $T_N$ of a finite system of length $L=Nd$ can be expressed through the following relation~\cite{LGP88,KTI97,IKM12},
\begin{equation}\label{eq:TN-R}
T_N=\frac{4}{2+(R^{(1)})^2+(R^{(2)})^2}\,.
\end{equation}
Here $R^{(1)}$ and $R^{(2)}$ are the radii of two independent trajectories at the time $n=N$ which start, respectively, from the points $(R^{(1)}_0,\theta^{(1)}_0)=(1,0)$ and $(R^{(2)}_0,\theta^{(2)}_0)=(1,\pi/2)$. In the present context, one can use the famous definition of the inverse localization length $L_{\mathrm{loc}}^{-1}$ via the transmittance,
\begin{equation}\label{eq:Lyap-lnTN}
\frac{d}{L_{\mathrm{loc}}}\equiv\lambda=-\lim_{N\rightarrow
\infty}\frac{1}{2N}\langle\ln T_N\rangle.
\end{equation}
As is known from the theory of disordered 1D systems, this definition of the Lyapunov exponent is equivalent to that considered above, see Eq.~\eqref{eq:Lloc-Lyap-def01}. In our numerical calculations we perform the averaging
$\langle\ldots\rangle$ over $5\times10^4$ different realizations of the disorder $v_n$ which allows one to reduce the fluctuations.

With the knowledge of the localization length, the mean value of $\ln T_N$ can be obtained due to the famous relation,
\begin{equation}\label{eq:mean-lnT}
\langle\ln T_N\rangle = - \frac{2 N d}{L_{\mathrm{loc}}},
\end{equation}
which can be rigorously derived for 1D weakly disordered models with continuous potentials, see for example, \cite{LGP88,IKM12}. It should be stressed that for such models this relation is valid for any ratio between the localization length $L_{\mathrm{loc}}$ and the sample size $L$, therefore, both in the ballistic regime (for $L_{\mathrm{loc}}\gg L$) and in the localized regime (for $L_{\mathrm{loc}}\ll L$).

In the case of finite one-dimensional continuous systems with weak random potential, the scattering problem was rigorously solved by various analytical methods. Our interest below is in the validity of the following rigorous expression for the moments of the transmittance $T_N$ (see, e.g., Refs.~\cite{IKM12,LGP88}) that can be readily adapted for a discrete model,
\begin{eqnarray}\label{eq:Ts}
&&\langle T_N^s\rangle=\sqrt{\frac{2}{\pi}}\,\frac{\exp(-\lambda N/2)}{(\lambda N)^{3/2}}
\int_0^\infty\frac{\alpha\exp(-\alpha^2/2\lambda N)d\alpha}{\cosh^{2s-1}\alpha}\nonumber\\
&&\times\int_0^\alpha d\beta\cosh^{2(s-1)}\beta,\quad
s=0,\pm1,\pm2,\pm3,\ldots\,.
\end{eqnarray}
Here $\lambda$ is the Lyapunov exponent, or the same, dimensionless inverse localization length $d/L_{\mathrm{loc}}\equiv\lambda$ (see details, e.g., in Ref.~\cite{IKM12}).

We would like to note that, strictly speaking, expression \eqref{eq:Ts} cannot be derived for the discrete models like our model with the wells and barriers. The reason is that for continuous potentials for which the expression has been derived, the resonances similar to those we are discussing here, are absent. It is known that for the standard tight-binding Anderson model the existence of the resonances do not allow to develop general analytical approach valid for any value of the energy of incident waves. The famous example is the band center for which the standard perturbation theory fails and one needs to use specific methods (see discussion and references in Ref.~\cite{IKM12}).

\begin{figure}[!!!t]
\includegraphics[width=8cm,height=5cm]{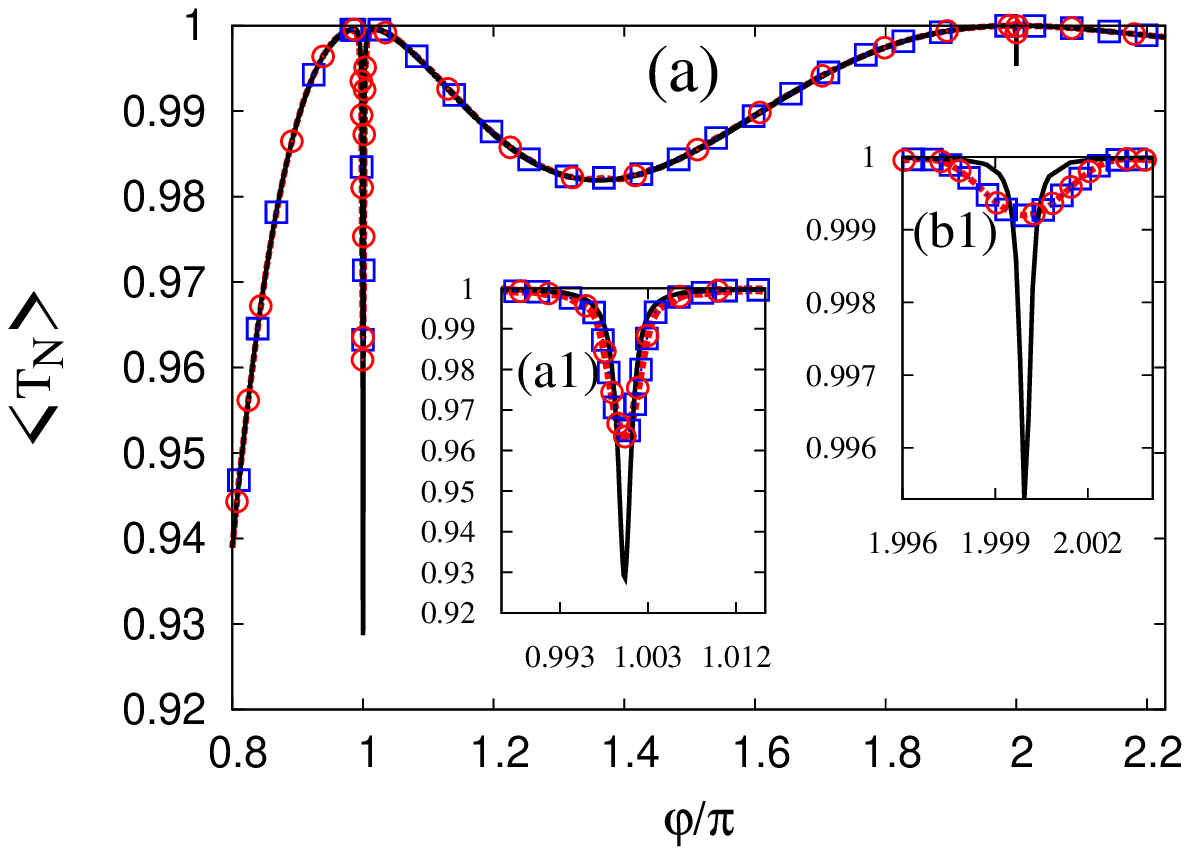}
\includegraphics[width=8cm,height=5cm]{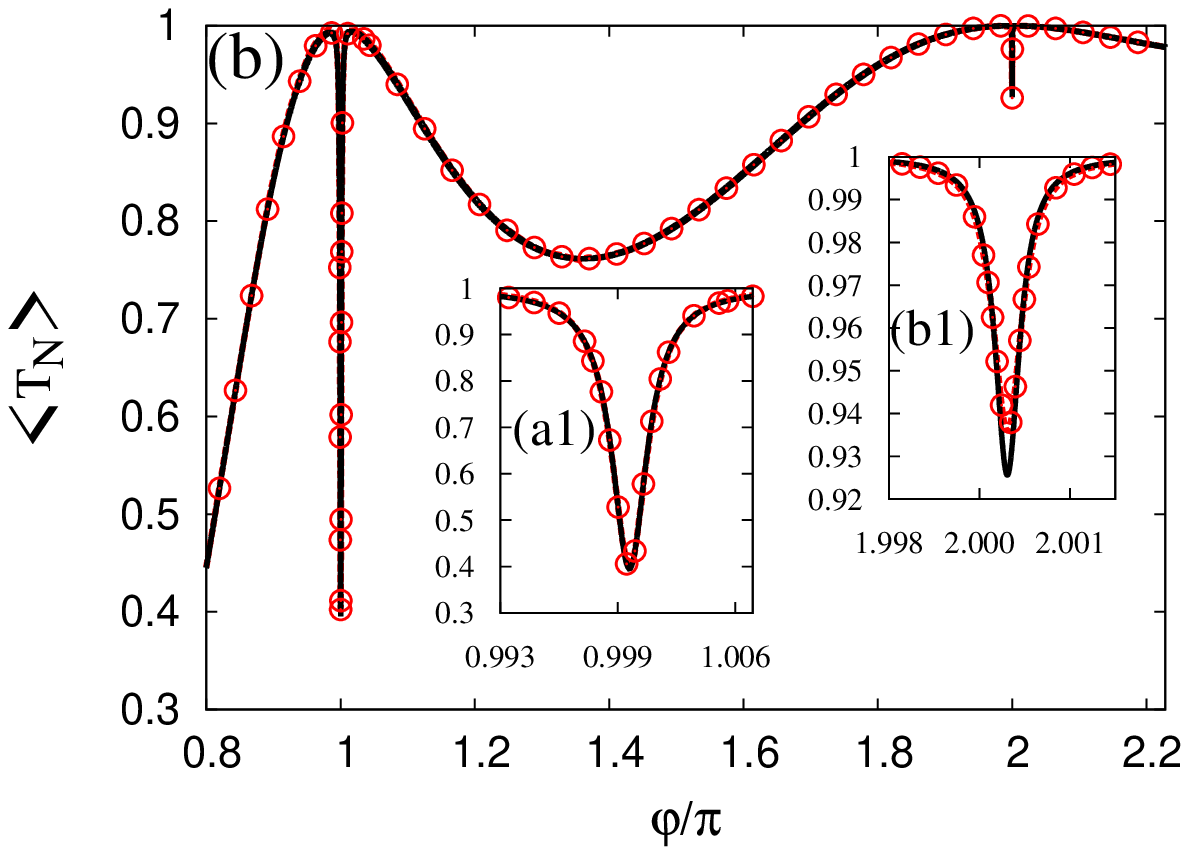}
\includegraphics[width=8cm,height=5cm]{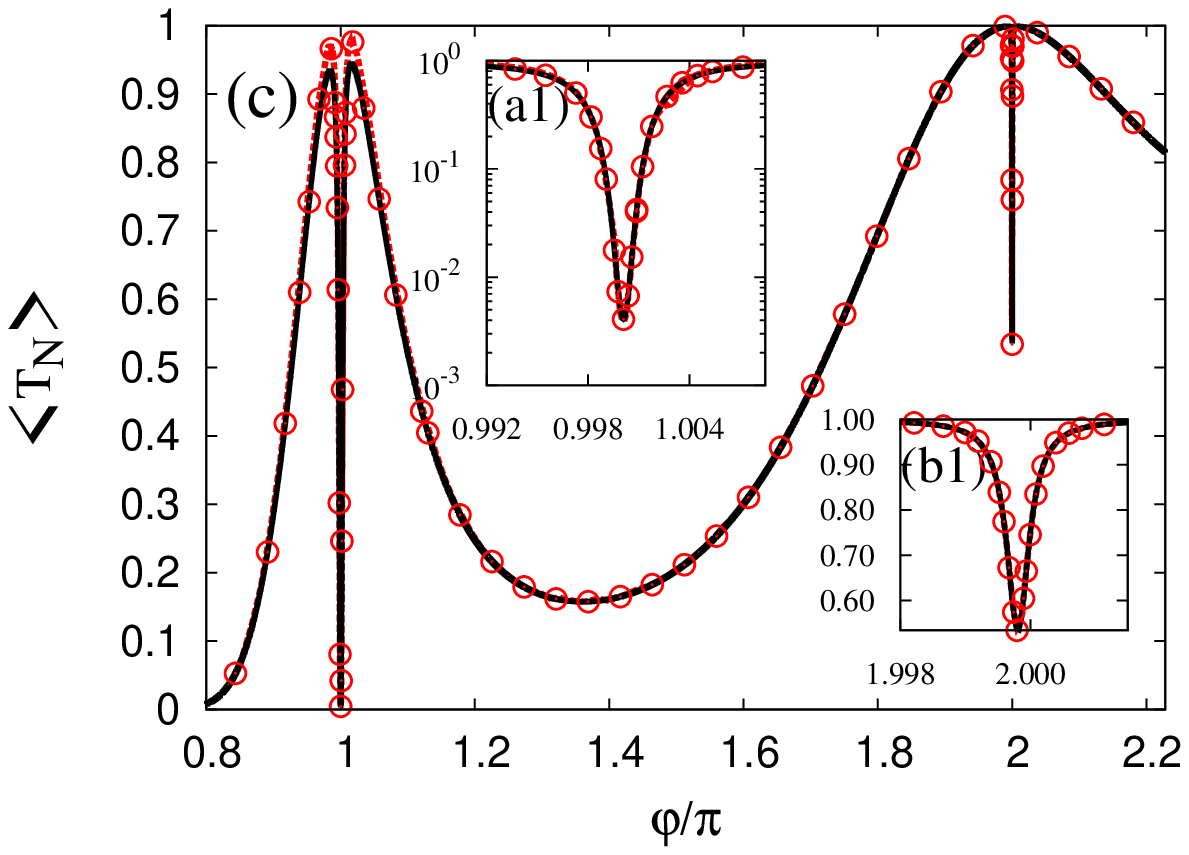}
\caption{\label{PRE-Fig07} (Color online) Average transmittance vs phase shift $\varphi$: (a) N=300, (b) N=5000, (c) N=50000. Continuous curves correspond to the analytical expression \eqref{eq:Ts} complemented by Eq.~\eqref{eq:Lyap-Fin}, circles represent the numerical simulation. Squares stand for the analytical Eq.~\eqref{eq:Ts} with $\lambda$ numerically computed from Eq.~\eqref{eq:Lyap-lnTN1}. Inset (a1) is a zoom of the resonant region at $\varphi=\pi$, inset (b1) is a zoom of the resonant region at $\varphi=2\pi$. The intensity of disorder is $\sigma^2=0.1$.}
\end{figure}

\begin{figure}[!!!t]
\includegraphics[width=8cm,height=5cm]{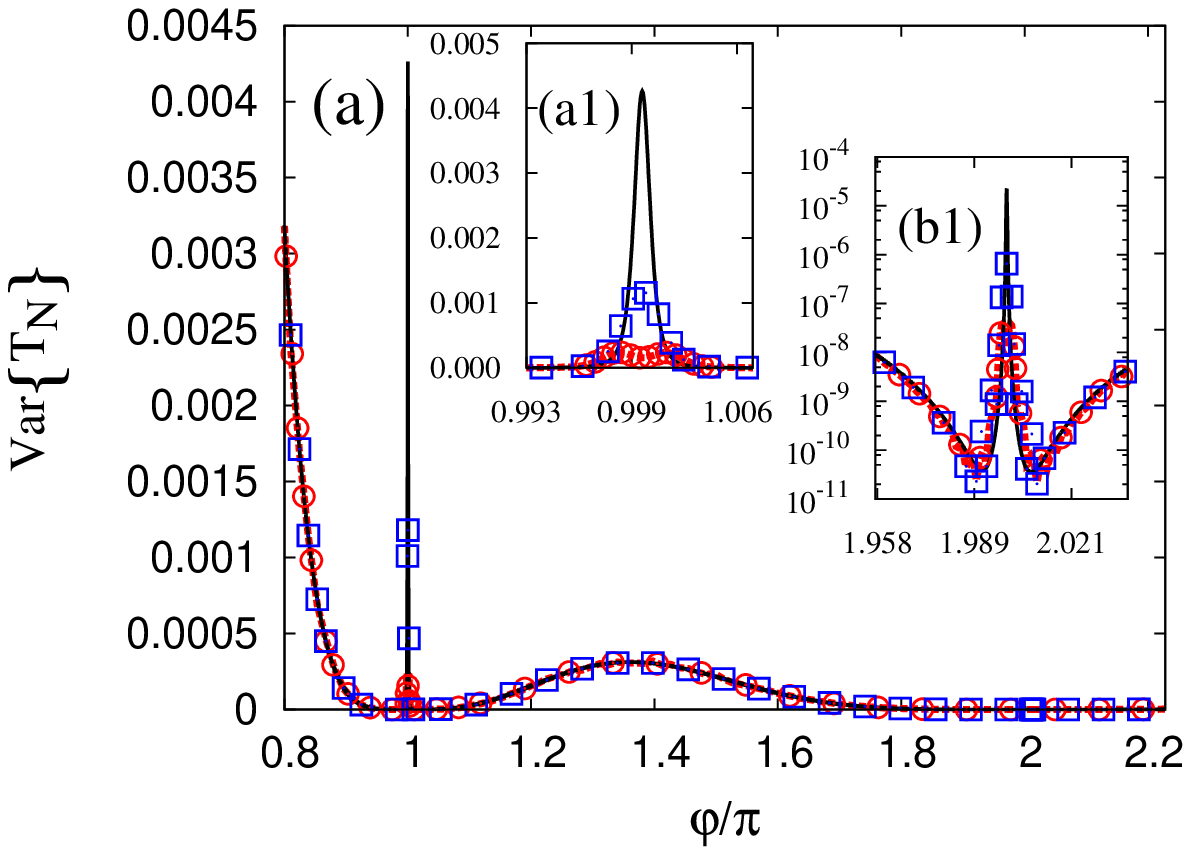}
\includegraphics[width=8cm,height=5cm]{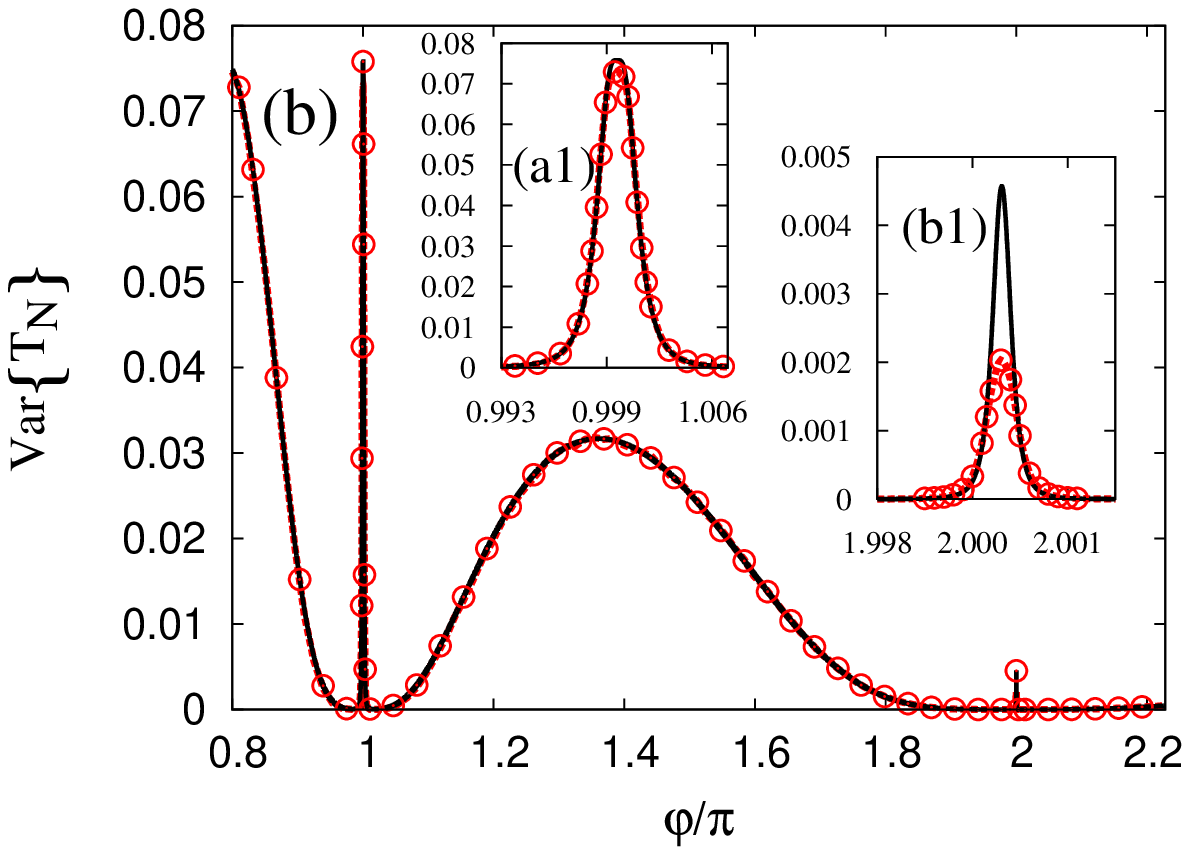}
\includegraphics[width=8cm,height=5cm]{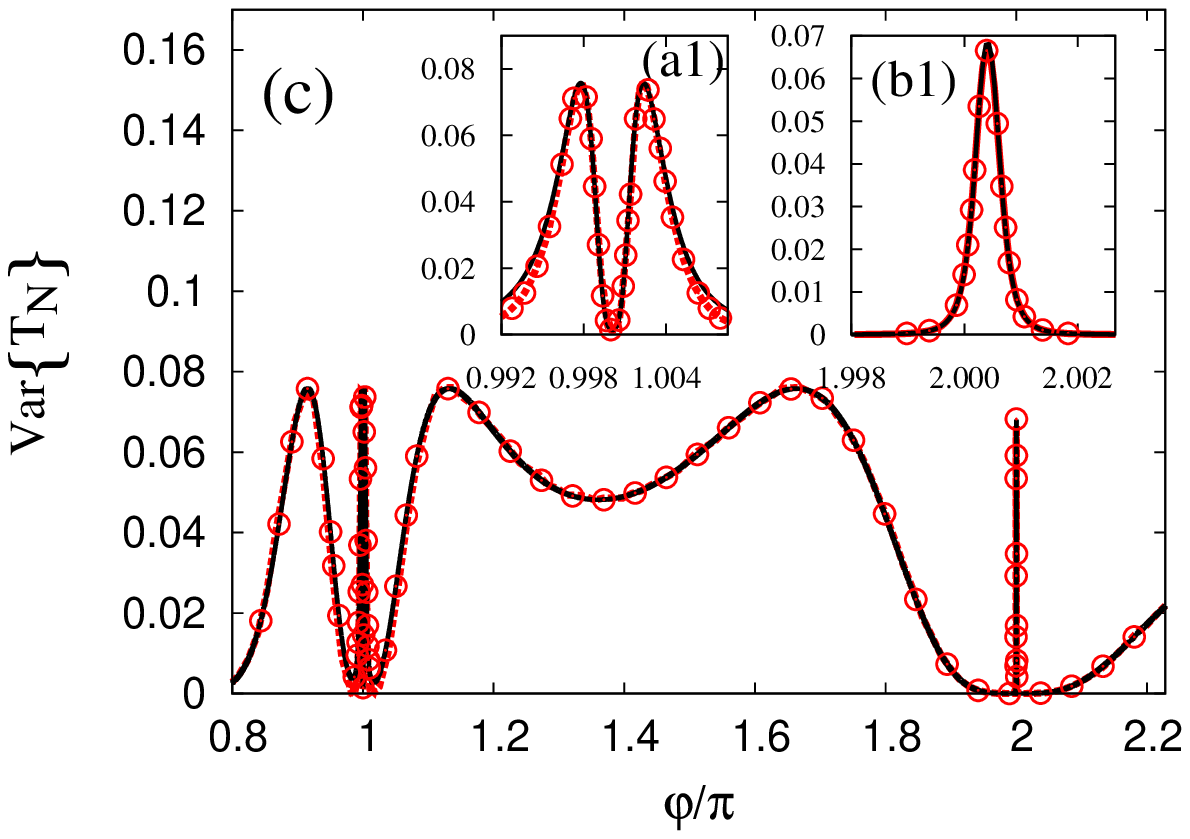}
\caption{\label{PRE-Fig08} (Color online) Transmittance variance vs $\varphi$. We use the same notations as in Fig.~\ref{PRE-Fig07}.}
\end{figure}

However, recently the expression \eqref{eq:Ts} has been tested for the Anderson model for non-resonant values of energy, and a perfect correspondence between the analytical predictions and numerical data has been manifested for two first moments of $T_N$ \cite{SIZC12}. Thus, our idea here is to explore the validity of the above expression for {\it both} non-resonant and resonant values of $\varphi$. In view of the results obtained in Ref.~\cite{SIZC12} we also expect that away from the resonances, where the phase distribution is flat, the formula \eqref{eq:Ts} gives correct
result. Indeed, the data in Figs.~\ref{PRE-Fig07} and \ref{PRE-Fig08} display an excellent agreement of the curves depicted with the use of Eq.~\eqref{eq:Ts} containing the Lyapunov exponent \eqref{eq:Lyap-flat}, with the numerical results calculated via Eq.~\eqref{eq:TN-R} in the regions of the phase shift $\varphi$ where phase distribution is flat.

On the other hand, our data have shown that in narrow regions of energy close to the resonances $\varphi=\pi$ and $\varphi=2\pi$, the expression \eqref{eq:Ts} completely ignores the presence of the resonances and gives incorrect results, provided the localization length is obtained by assuming the phase distribution is flat. Thus, our key point is to explore whether the same expression \eqref{eq:Ts}, however, with the correct Lyapunov exponent can serve both for non-resonant and resonant values of $\varphi$.

The idea to combine the standard expression \eqref{eq:Ts} with the Lyapunov exponent \eqref{eq:Lyap-Fin} valid both in the non-resonant and resonant regions turns out to be very fruitful. Indeed, the data in Figs.~\ref{PRE-Fig07}(b,c) and \ref{PRE-Fig08}(b,c) demonstrate an excellent agreement with our analytical predictions not only for the mean value of $T_N$ but also for the variance $\mbox{Var}\{T_N\}\equiv\langle T_N^2\rangle-\langle T_N\rangle^2$.

Note, however, that our approach does not work if the value of $N$ is not large enough. This fact is clearly seen in Figs.~\ref{PRE-Fig07}(a) and \ref{PRE-Fig08}(a) where $N=300$. The estimate of the Lyapunov exponent $\lambda$ for the chosen strength of disorder $\sigma^2=0.1$ shows that for both resonances the localization length $L_{\mathrm{loc}}$ is much larger than the system size $Nd$, this corresponds to the ballistic regime. As one can see, a good correspondence between the data and our analytical approach occurs in a strongly localized regime, Figs.~\ref{PRE-Fig07}(c) and
\ref{PRE-Fig08}(c), and in the intermediate regime where the localization length $L_{\mathrm{loc}}$ is of the order of $Nd$.

Again, we have to recall that the analytical expression \eqref{eq:Ts} works well for any ratio between the localization length and system size, provided the disorder is described by continuous potentials for which there are no resonance effects. The failure of our approach in the ballistic regime is due to the non-stationarity of the phase distribution as we explain below.

A closer inspection of the equation \eqref{eq:thetaMap-appr} describing the evolution of phase $\theta$ shows that at the resonances the filling of the whole range $[0,2\pi]$ by the phase is due to the terms containing the disorder, and not due to the constant drift due to non-resonant values of $\varphi$. Therefore, the length $N_{cr}$ for the emergence of a stationary distribution for $\theta$ can be very large, in contrast with what happens out of resonances. Indeed, a rough estimate of this critical length $N_{cr}$ gives $N_{cr}\approx8\varphi^2/\sigma^2$. Thus, we have
$N_{cr}\approx800$ and $N_{cr}\approx3600$ for the resonances $\varphi=\pi$ and $\varphi=2\pi$, respectively. These estimates explain the discrepancy which can be seen in Figs.~\ref{PRE-Fig07}(a,b) and \ref{PRE-Fig08}(a,b). According to Eq.~\eqref{eq:Lyap-Fin} the localization length at the resonances is $L_{\mathrm{loc}}/d=8\varphi^4/\sigma^2$. As one can see, the ratio between $L_{\mathrm{loc}}$ and $N_{cr}d$ is $\varphi^2$, the estimate which gives an additional information about the role of resonances in our model.

%
%
\begin{figure}[!ht]
\includegraphics[width=8.0cm,height=5cm]{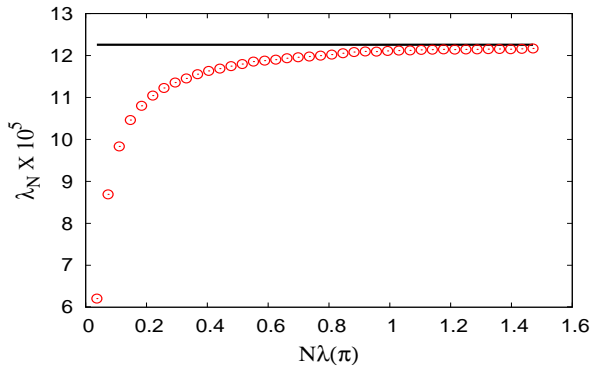}
\caption{\label{PRE-Fig09} (Color online) Resonant size-dependent Lyapunov exponent $\lambda_N$ at $\varphi=\pi$ vs system size $N$. Circles show $\lambda_N$ numerically calculated from Eqs.~\eqref{eq:TN-R}, \eqref{eq:Lyap-lnTN1}, straight line presents $\lambda(\pi)=\sigma^2/8\varphi^4$ defined by analytical Eq.~\eqref{eq:Lyap-Fin}. The system length $N$ is rescaled by $\lambda(\pi)$. The intensity of disorder is $\sigma^2=0.1$.}
\end{figure}

As the next step towards a better agreement between the analytical description and numerical data, we can suggest to use the size-dependent Lyapunov exponent $\lambda_N$ defined as follows,
\begin{equation}\label{eq:Lyap-lnTN1}
\lambda_N=-\frac{1}{2N}\langle\ln T_N\rangle.
\end{equation}
Then, one can try to use it in the integral formula~\eqref{eq:Ts}, instead of the stationary Lyapunov exponent $\lambda$. It turns out that in this case one can get a quite good agreement with the numerical simulations, at least for the first two cumulants of the transmittance as Figs.~\ref{PRE-Fig07}(a) and \ref{PRE-Fig08}(a) show.

As one can see, the analysis of the ballistic regime ($\lambda N\ll1$) in the vicinity of the resonances where $\varphi\approx j\pi$ requires two definitions, Eqs.~\eqref{eq:Lloc-Lyap-def01} and \eqref{eq:Lyap-lnTN1}, for the Lyapunov exponent. The first (standard) definition \eqref{eq:Lloc-Lyap-def01} is given for an infinite system, therefore, the stationary $\theta$-distribution is always achieved. On the other hand, Eq.~\eqref{eq:Lyap-lnTN1} is the prelimit counterpart of the first one, and, therefore, depends on the system size $N$. Thus, Eq.~\eqref{eq:Lyap-lnTN1} automatically takes into account an actual phase distribution and provides quite good results even in the resonant ballistic regime, where the phase distribution is non-stationary. Fig.~\ref{PRE-Fig09} demonstrates that the size-dependent Lyapunov exponent \eqref{eq:Lyap-lnTN1} is saturated and becomes equivalent to the size-independent one \eqref{eq:Lloc-Lyap-def01} at $\lambda N\gtrsim1$. Consequently, at this system length the stationarity of $\theta$-distribution is successfully reached, and the analytical expression \eqref{eq:Lyap-Fin} is valid. Indeed, the numerical simulations originated from Eqs.~\eqref{eq:TN-R}, \eqref{eq:Lyap-lnTN1} are in excellent agreement with analytical equations \eqref{eq:Ts} and \eqref{eq:Lyap-Fin}. This fact is clearly seen in Figs.~\ref{PRE-Fig07}(b,c) and \ref{PRE-Fig08}(b,c).

Our results demonstrate that the formulas \eqref{eq:Ts}, \eqref{eq:Lyap-Fin} provide quite good analytical description of the transport properties in comparison with numerical simulation even at the resonant energies. This may suggest that the hypothesis of single parameter scaling (SPS) is correct provided the value of the localization length takes into account the non-flat distribution of phases $\theta$ in the vicinity of resonances. It should be however stressed that there are, at least, two different definitions of the SPS. The first one, which is a trivial consequence of the expression
\eqref{eq:Ts}, is that all transport properties depend on the ratio between the localization length and the sample size. Another definition is originated from the analysis of tight-binding models of the Anderson type and various Kronig-Penney models (see discussion in Ref.~\cite{IKM12}). Since for such models the resonances are unavoidable, the rigorous analysis in the general form is absent. For this reason one uses another definition of the SPS according to which all properties of the transport depend on the mean logarithm of transmittance, $\langle\ln T_N\rangle$. This is correct, e.g., for the Gaussian distribution of $\ln T_N$ (log-normal distribution) if the ratio
\begin{equation}\label{eq:R2}
R_2\equiv-\frac{\mbox{Var}\{\ln T_N\}}{\langle\ln T_N\rangle}=
\frac{\langle\ln T_N\rangle^2-\langle\ln^2 T_N\rangle}{\langle\ln T_N\rangle}.
\end{equation}
is assumed to be constant. The latter occurs in strong localization regime and when random phase hypothesis holds true. Under such circumstances $R_2$ equals $2$, and this result is often used as a proof or disproof of the SPS (see, e.g., \cite{ST03a,DELA03} and references therein). However, one has to bear in mind that random phase hypothesis is not a necessary condition for the validity of the SPS \cite{DLA00}.

\begin{figure}[!ht]
\includegraphics[width=8.0cm,height=5cm]{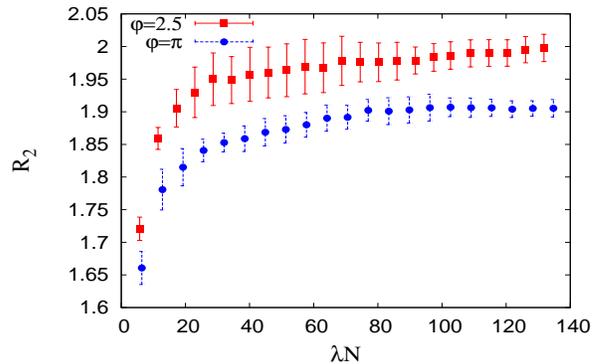}
\caption{\label{PRE-Fig10} (Color online) The parameter $R_2$ as a function of the system size rescaled by the inverse localization length $\lambda$ for a disorder intensity of $\sigma^2=1/2$. Circles represent numerical data at the resonant phase shift $\varphi=\pi$ and $\lambda=\lambda(\pi)$, while squares are the numerical result for the non-resonant value $\varphi=2.5$ with $\lambda=\lambda(2.5)$ . The averaging $\langle\ldots\rangle$ was performed over $10^4$ different realizations of the disorder. To obtain the error bars an ensemble of $100$ different random values of $R_2$ was used.}
\end{figure}

In view of this common approach, we have performed a careful numerical calculations of the parameter $R_2$ in a wide range of the sample size $N$. The results shown in Fig.~\ref{PRE-Fig10} turn out to be quite unexpected. Namely, in spite of a good description of the first and second moments of $T_N$ by the expression \eqref{eq:Ts} with the correct Lyapunov exponent \eqref{eq:Lyap-Fin}, we have found the failure of the SPS when exploring the variance of $\ln T_N$. This is in a strong contrast with the result according to which if Eq.~\eqref{eq:Ts} for the moments of $T_N$ is valid (and, therefore, the whole distribution of $T_N$ is known), the value of the parameter $R_2$ has to be 2, as is predicted by the theory for the continuous systems (see, e.g., \cite{IKM12} and references therein). However, our data clearly demonstrate that $R_2=1.91\pm0.02$ at the resonant phase shift $\varphi=\pi$. Whereas $R_2=1.99\pm0.035$ for the non-resonant phase shift $\varphi=2.5$ in accordance with our expectation. Due to these results, one may conclude that the SPS is not valid for resonant energies, however, more extensive studies, both the analytical and numerical ones, are required in order to resolve an apparent paradox demonstrated by the data.

\begin{acknowledgments}
The authors are thankful to Pier Mello for fruitful discussions. We acknowledge support form the SEP-CONACYT (M\'exico) under grant No. CB-2011-01-166382, VIEP-BUAP grant MEBJ-EXC12-G, and PIFCA BUAP-CA-169.
\end{acknowledgments}



\end{document}